\documentclass[twocolumn,preprintnumbers,showpacs,amsmath,amssymb,prb]{revtex4}

\usepackage{graphicx}
\usepackage{dcolumn}
\usepackage{bm}
\usepackage{amssymb}
\usepackage{chemist}

\begin{document}
\title{Optical valleytronics in gapped graphene}

\author{K. Dini$^{1,2,3}$, I. V. Iorsh$^2$, A. Bogdanov$^2$
and I. A. Shelykh$^{1,2}$}
\address{$^1$ Science Institute,
University of Iceland, Dunhagi 3, IS-107, Reykjavik, Iceland}
\address{ $^2$ ITMO University, Saint Petersburg 197101, Russia}
\address{ $^3$ Division of Physics and Applied Physics, Nanyang Technological University 637371, Singapore}

\begin{abstract}
We propose a scheme to trap and filter electrons, valley dependently, on a scale beyond the diffraction limit, in a gapped Dirac system using a circularly polarized light beam and a micro- scale metallic resonator. The main mechanism allowing the selection is the valley dependent break of the time reversal symmetry. Indeed in one valley, the light partially closes the already existing gap while it opens it in the other one. This difference in the band structure close to the Dirac points, induces a change in the dynamics of the electrons, leading to a valley router behavior of the system. 
\end{abstract}

\maketitle
\section{Introduction}

The discovery of graphene, a monolayer of carbon atoms with low energy linear dispersion, and its unique electronic properties still continue to create an increasing interest of the scientific community \cite{Geim_07,CastroNeto_09,DasSarma_11}, and initiated a large amount of studies on a new class of nanostructures, the \textit{Dirac materials}. Graphene is characterized by the gapless energy spectrum, which makes difficult its application in modern electronic fields. Therefore many efforts have been dedicated towards the fabrication of gapped Dirac materials. The spectrum of those materials is parabolic near the band edge (located at the so-called Dirac points) but becomes linear far from it. Therefore electronic properties of such materials strongly depend on the value of the band gap and, consequently, they are suitable for potential nanoelectronic applications  \cite{Lensky_15,Zhu_17,Kundu_16}. Like in bare gapless graphene, the confinement of Dirac electrons in such systems, and the wide range of potential practical applications, are compromised by their ability to perfectly transmit through arbitrary potentials barriers at normal incidence. This effect is well known for Dirac particles as the Klein paradox \cite{Katnelson_06}. 

Valley transport in graphene has recently become a very active field of research since it is expected that the valley degree of freedom can play the same role as the electron spin in information processing \cite{Pesin_12,Wang_17,Rycerz_07}. Today this approach is referred as \textit{valleytronics} and it has many similarities with spintronics. In graphene like systems, the valley degree of freedom exists due to the fact that there are two inequivalent edges of the Brillouin zone in the honeycomb lattice labbelled as K and K'. Due to the large distance between those two valleys in the reciprocal space, only scatterers with a range smaller that the lattice constant can induce intervalley scattering, for that reason it is usually considered very weak and can be neglected in clean samples. Therefore, the valley index is a quantum number that can be considered as conserved for electron transport.  A lot of proposals in the literature were put forward to generate valley-polarized currents by using graphene nanoribbons \cite{Akhmerov_08,Rycerz_07}, electromagnetic or optical field \cite{Golub_11,Zeng_12} and lattice strain \cite{Wu_16,Fujita_10}.
 
In the present letter, we propose a way to achieve valley dependent optical trapping of Dirac electrons in gapped Dirac systems and to create a valley router.  In the framework of condensed matter physics, the basis for optical trapping is provided by the possibility to locally modify the energy spectrum of the particles by strong coupling to high frequency electromagnetic field resulting in the dynamic Stark effect. Strong modifications of the transport properties in the regime of strong light-matter couplig have been recently reported for semiconductor quantum well \cite{Wagner_10,Teich_13,Morina_15,Pervishko_15,Dini_16}, carbon nanostructures \cite{Oka_09,DalLago_17,Iorsh_17,Kristinsson_16,Kibis_16,Monroy_16} and topological insulators \cite{Yudin_16,Torres_14,Usaj_14}. 

Particularly, it has been shown that in gapped Dirac systems strongly coupled to circularly polarized light the value of the band gap $\Delta_g$ is either increased or decreased valley dependently \cite{Kibis_17}. This modification depends on the intensity $I$ and the frequency $\omega$ of the driving field. Therefore, this mechanism offers a reversible way to control the electronic properties of the system. Indeed, let us consider the case of a non-homogeneous intensity of the driving field i.e. a simple Gaussian shaped beam (see figure 1). In this case, the band gap is  modified in the vicinity of the beam and it will be unchanged away from it. Therefore, an electron located inside the illuminated area, will not be able to escape since there is no available propagating states. This mechanism has been described before for bare graphene \cite{Morina_17} and is similar to the electron confinement in semiconductor heterostructures when a layer of narrow band semiconductor is sandwiched between wide gap materials with the only difference being that the band mismatch in this case is produced  all optically. In gapped Dirac systems, the valley dependence of the band gap has to be taken into account, since the gap increases in one valley and decreases in the other one, the shined area will respectively be a forbidden area for low energy electrons or a trapping one. Moreover, since the energy spectrum is locally changed, electrons propagating towards the illuminated area feel it as a potential barrier, and therefore valley dependent scattering will occur. In this letter we propose and describe a system, based on this principle, which traps electrons valley dependently and filters electrons moving towards the shinned area. 

\begin{figure}
\centering
\includegraphics[width=0.5\textwidth]{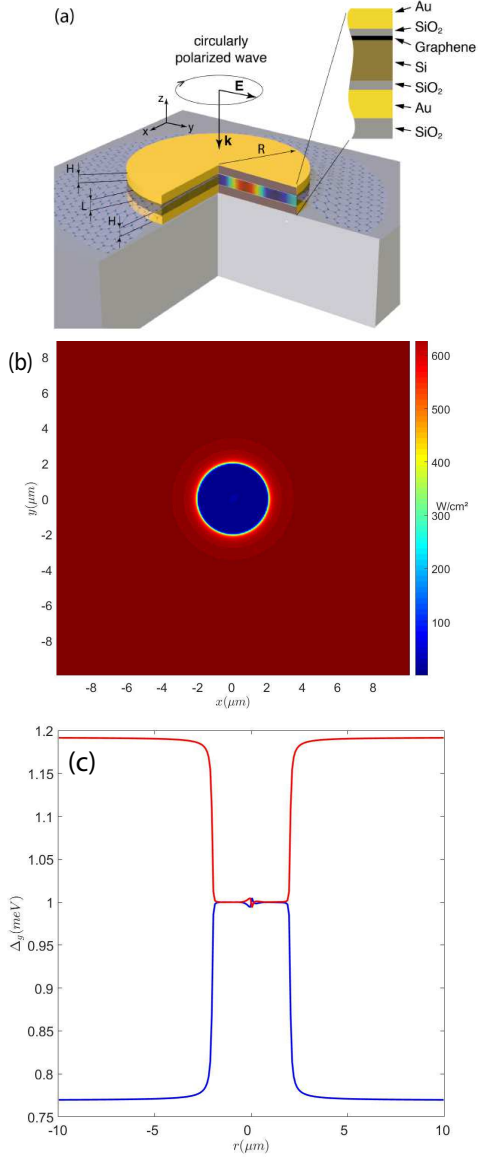}
\caption{(a) Sketch of the considered system. The presence of a micro- scale metallic resonator allows to focus high frequancy electromagnetic field at subwavelength scale  (b) Spacial profile of the intensity of the field (c) Spacial dependence of the gap on the radial distance form the center of the beam in K (red) valley and K' (blue) valley.}
\label{Dependence}
\end{figure}

\begin{figure*}
\centering
\includegraphics[width=\textwidth]{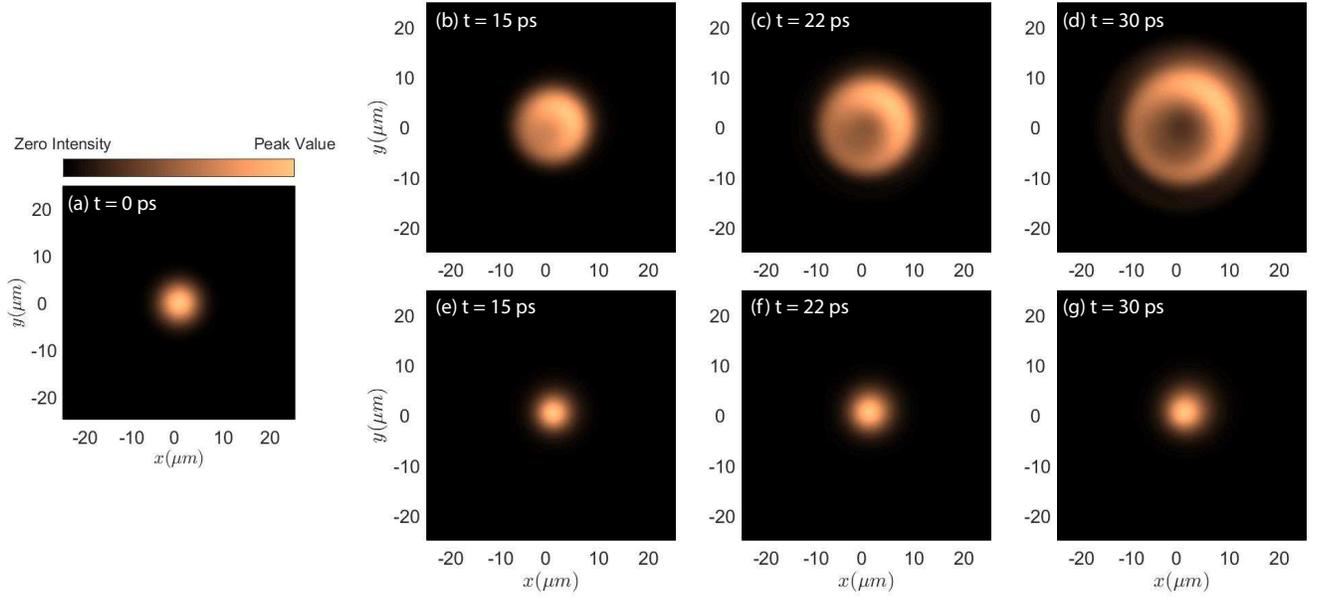}
\caption{Plots of the total electron density, $\rho=|\psi_A|^2 + |\psi_B|^2$, for several values of the evolution time
indicated in each subfigure. Panel \textbf{(a)} corresponds to the initial distribution at $t=0$. Panels \textbf{(b)-(d)} correspond to the dynamics in the K valley. Panels 
 \textbf{(e)-(g)} correspond to the dynamics in the K' valley.  One clearly sees that while electrons located in the K valley escape the trap, those located in K' valley remain confined in it.}
\label{fig:IntensityTrap}
\end{figure*}

\begin{figure*}
\centering
\includegraphics[width=\textwidth]{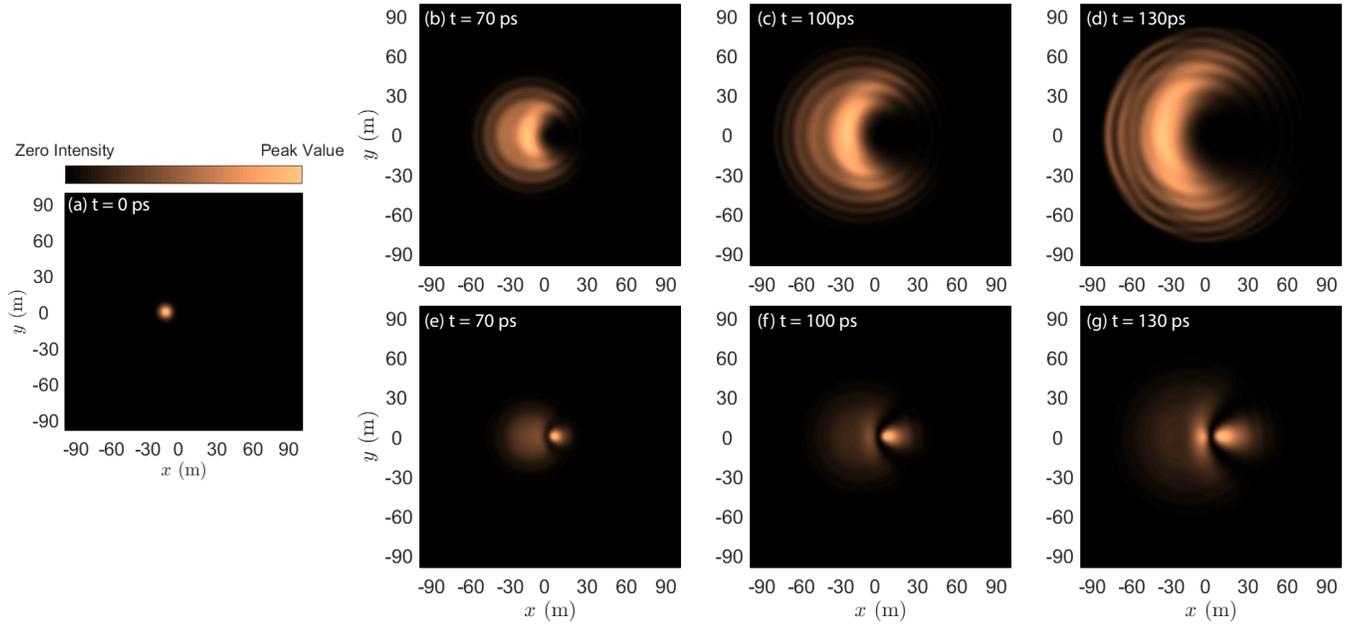}
\caption{Plots of the total electron density, $\rho=|\psi_A|^2 + |\psi_B|^2$, for several values of the evolution time
indicated in each subfigure. Panel \textbf{(a)} corresponds to the initial distribution at $t=0$. Panels \textbf{(b)-(d)} correspond to the dynamics in the K valley. Panels 
 \textbf{(e)-(g)} correspond to the dynamics in the K' valley. One clearly sees different scattering patterns for electrons in K and K' valleys.}
\label{fig:IntensityFilter}
\end{figure*}

\section{Model}

The bare system we consider consists of an infinite single layer of graphene which lies in the plane $\mathbf r = (x,y)$ at $z=0$, grown between two 3~nm thick layers of \chemform{SiO_2}. We locally add a circular metallic resonator with a $R=2 \ \mu m$ radius, centered at $(x,y)= 0$, the top \chemform{SiO_2} layer is directly stuck below the top gold plate while on the other side a thick silicon layer is grown [see Fig \ref{Dependence} (a)]. The resonator is added in order to break through the diffraction limit and to obtain micro-meter scale trap/filter in the infra-red range. The system is coupled to an electromagnetic wave propagating along the z-axis. The influence of the resonator on the effective distribution of the electromagnetic field is modeled numerically. The in-plane component of the field is displayed in Fig \ref{Dependence} (b).

In order for the field to be considered as a pure dressing one, the absorption coefficient in the vicinity of the Dirac point should be close to zero. Therefore, the frequency of the electromagnetic field must be chose to be high enough to satisfy the condition $\omega\tau\gg1$ where $\tau$ is a characteristic relaxation time.
The low- energy Hamiltonian of the system reads: 
\begin{equation}\label{H0}
\hat{{\cal H}}(t)=\hbar v_F\left[\xi \hat{\sigma}_x \left(k_x+\frac{eA_x}{\hbar}\right) + \hat{\sigma}_y \left(k_y+\frac{eA_y}{\hbar}\right)\right]+\frac{\Delta_g}{2}\hat{\sigma}_z  
\end{equation}
where $v_F$ is the Fermi velocity, $\xi=\pm1$ the valley index and $\sigma_i$, $i = x,y$, are the Pauli matrices. The description of the electron-photon coupling can be done via the minimal coupling approach performing the following canonical transformation:
$k_{x,y} \rightarrow k_{x,y} + (e/\hbar) A_{x,y} $ where $A_{x,y}$  corresponds to the vector potential of the dressing field.  In order to open a band gap at the Dirac points, the dressing field must be chosen to be circularly polarized: 
\begin{equation}
\label{VectorPotential}
A_x=\frac{E_{x}(r,\theta)}{\omega}\cos\left(\omega t \right),\\
A_y=\frac{E_{y}(r,\theta)}{\omega}\sin\left(\omega t \right),
\end{equation}
where $\omega$ and $E_{x,y}$ are  frequency and amplitudes of the dressing field.

For the purpose of studying the system in the stationary regime, we need to reduce Eq~\ref{H0} to a time-independent effective Hamiltonain. To do so we here choose to use the Floquet theory of periodically driven quantum systems 
\cite{Hanggi_98,Kohler_05,Bukov_15,Goldman_14}.
The main steps are as follows. The time-dependent Hamiltonian can be 
expressed as : 

\begin{align}
\label{SeparationHarmonics}
\hat{\cal H}\left(r,t\right)= \hat{ \cal H}_0+ 
\hat{V} \exp\left(i \omega t \right)+ 
\hat{V}^\dagger \exp\left(-i \omega t \right)
\end{align}

where
\begin{align}\label{Harmonics}
\hat{{\cal H}}_0(t)&=\hbar v_F\left(\xi \hat{\sigma}_x k_x+ \hat{\sigma}_y k_y\right)+ \frac{\Delta_g}{2} \hat{\sigma}_z,\\
\hat{V}\left(r\right)&=  \frac{e v_F}{2 \omega}\left(\xi E_{x} \hat{\sigma}_x - 
iE_{y}\hat{\sigma}_y\right).
\end{align}

Since the frequency $\omega$ is assumed to be high compared to all characteristic frequencies of the system, the electron dynamics is not able to follow the fast time oscillations of the vector potential, and the effective time-independent Hamiltonian can be obtained by Floquet-Magnus expansion
\cite{Goldman_14,Rahav_03,Eckardt_15} in powers of $\omega^{-1}$. Restricting ourselves to the first three terms in the infinite series we get:
\begin{equation}
\nonumber\hat{{\cal H}}_{\mathrm{eff}}\approx\hat{{\cal H}}_0+\frac{\left[\hat{V},\hat{V}^\dagger\right]}{\hbar\omega}+
\frac{ \left[\left[\hat{V},\hat{{\cal H}}_0\right],	\hat{V}^\dagger\right]+
\mathrm{H.c.}}{2(\hbar \omega)^2}
\label{Effective}
\end{equation}

Using this expression, one can find out the renormalization of the band parameters of the system \cite{Kibis_17}, the modified band-gap can be expressed as: 
\begin{equation}
\label{GapG}
\tilde{\Delta}_g = \Delta_g \left(1 - \left(\Omega_x^2 + \Omega_y^2 \right)\right)- 2 \xi \hbar \omega\Omega_x \Omega_y
\end{equation}
where 

\begin{equation}
\Omega_{x,y}= \frac{v_F e E_{x,y}}{\hbar \omega^2}
\end{equation}

This term is responsible for the valley dependent trapping of electrons with energy below the bare gap value and for the valley dependent scattering of electrons. The position-dependent renormalized Fermi velocity $\tilde{v}_{x,y} = v_F ( 1 - \Omega_{x,y}^2)$. One should note that since the distribution of the field is space dependent, non-commutative terms appear in the effective Hamiltonian due to the canonical commutation relation. Those terms are small with respect to the other terms but are retained in the following simulations in order to keep the effective Hamiltonian Hermitian.

It should be noted that Eq.(\ref{Effective}) is derived under the condition $\hbar \omega \gg \Delta_g$. The gap in such system can be tuned in the broad range $\Delta_g = 1 - 60$ meV \cite{Sachs_11,Jung_15}. Therefore, assuming the gap to be of meV scale, and the field to be in the far infrared range, we can easily satisfy this condition. 

\subsection{Results and discussion}

In order to demonstrate the valley dependence of the electron dynamic and its consequences on the optical trapping of electrons in gapped Dirac systems, we first study numerically the behavior of an electronic wave packet injected at $t=0$ in the center of the resonator. The dressing field has the following characteristics; intensity $I =$~300~W/cm$^2$, frequency $\omega =33$~THz. The initial average wave vector of the packet is null and its width is $d = 4 \ \mu$m. The simulation is run for K and K' independently i.e. the electron wave packet is injected in one valley at a time. The conservation of the global intensity has been checked and is verified up to $ 10^{-6} \% $. The results are shown in FIG. \ref{fig:IntensityTrap}. From this simulation one can conclude that, in the valley where the gap in the area between the metallic plates is lower than elsewhere, described by panes (e) to (g), electrons cannot find any possible state to propagate to, and therefore stay in the vicinity of $r=0$. This mechanism has already been described in recent studies \cite{Morina_17}. In the other valley, described by panels (b) to (d), when the gap is smaller, propagating states with not zero k vectors exist, and depending on the energy at which the electron is injected, not zero velocity can be observed. Also, this valley dependence can be optically controlled using the polarization of the field. Namely, switching form a clockwise to a counterclockwise circular polarization will switch the effect to its opposite in each valley \cite{Kibis_17}. 

In order to study the interaction between a propagating electron wave packet and the light induced valley and space dependent modification of the gap, it is now introduced at $ x = -15 \mu$m , $ y = 0 $  with the same characteristics as the previous study. The results are shown in Fig.\ref{fig:IntensityFilter}. From this simulation one can conclude that in both valleys, the dynamic of the propagation wave packet is strongly affected by the modification of the gap. In the K valley, described by panels (e) to (g), where the gap is decreased, part of the total intensity is transmitted and will continue to propagate to the right. The appearance of a "fan" pattern is due to the radial symmetry of the light induced effective potential. In the K' valley, the wave packet is scattered in every direction but the Ox one. Simulations with different initial energies of the wavepacket have been run, this effect subsists up to an initial energy of the packet close to the value of the modified band gap. As in the case of the valley dependent trapping, the valley can be switched by changing the polarization. 

\subsection{Conclusion}

In conclusion, in this work we have proposed a scheme that can both behave as a valley dependent trap for electrons and as a valley router, on a scale lower than the one usually allowed by the diffraction limit. The basis on which this results lies on is the all optical valley dependent modification of the band gap. This modification of the band parameters changes the position of the electronic propagating states and therefore electrons can be either trapped, reflected or transmitted. Numerical simulations supporting the prediction of the trap and filter behavior of the system are provided. This findings open the routes to the all-optical manipulation of the valley transport in 2d materials with the subwavelength resolution. 

\subsection{Acknowledgments }

The work was partially supported by the Russian Foundation for Basic Research (project 17-02-00053), RISE Program (project CoExAN) and Ministry of Education and Science of Russian Federation, Projects  No. 3.2614.2017/4.6 and No. 14.Y26.31.0015) and Government of Russian Federation (Grant No. 08-08).


\begin{thebibliography}{99}

\bibitem{Geim_07}
A. K. Geim and K. S. Novoselov, Nature Materials 6, 183–191 (2007).

\bibitem{CastroNeto_09}
A. H. Castro Neto, F. Guinea, N. M. R. Peres, K. S. Novoselov and A.
K. Geim, Rev. Mod. Phys. {\bf 81}, 109 (2009).

\bibitem{DasSarma_11}
S. Das Sarma, S. Adam, E. H. Hwang and E. Rossi, Rev. Mod. Phys.
{\bf 83}, 407 (2011).

\bibitem{Lensky_15}
Y.D. Lensky, J.C.W. Song, P. Samutpraphoot and L. S. Levitov,
Phys. Rev. Lett. \textbf{114}, 256601 (2015).

	
\bibitem{Zhu_17}	
M. J. Zhu, A. V. Kretinin, M. D. Thompson, D. A. Bandurin, S. Hu, G. L. Yu, J. Birkbeck, A. Mishchenko, I. J. Vera-Marun, K. Watanabe, T. Taniguchi, M. Polini, J. R. Prance, K. S. Novoselov, A. K. Geim and M. Ben Shalom, Nature Communications \textbf{8}, 14552 (2017). 

\bibitem{Kundu_16}
A. Kundu, H.A. Fertig and B. Seradjeh, 
Phys. Rev. Lett. \textbf{116}, 016802 (2016).

\bibitem{Katnelson_06}
M. I. Katsnelson, K. S. Novoselov and A. K. Geim, Nature Physics \textbf{2}, 620–625 (2006).

\bibitem{Pesin_12}
D. Pesin and A. H. MacDonald, Nature Materials \textbf{11}, 409–416 (2012).

\bibitem{Wang_17}
J. J. Wang, S. Liu, J. Wang and J. Liu, Sci. Rep. \textbf{7}, 10236 (2017).

\bibitem{Rycerz_07}
A. Rycerz, J. Tworzydlo and C. W. J. Beenakker, Nature Physics \textbf{3}, 172–175 (2007).

\bibitem{Akhmerov_08}
A. R. Akhmerov, J. H. Bardarson, A. Rycerz and C. W. J. Beenakker,
Phys. Rev. B \textbf{77}, 205416 (2008).

\bibitem{Golub_11}
L. E. Golub, S. A. Tarasenko, M. V. Entin and L. I. Magarill, Phys. Rev. B \textbf{84}, 195408 (2011).

\bibitem{Zeng_12}
H. Zeng, J. Dai, W. Yao, D. Xiao and X. Cui, Nat. Nanotechnol. \textbf{7}, 490 (2012).

\bibitem{Wu_16}
Q. Wu, Z. Liu, A. Chen, X. Xiao and Z. Liu, Sci. Rep \textbf{6}, 21590 (2016).

\bibitem{Fujita_10}
T. Fujita, M. B. A. Jalil1, and S. G. Tan, Appl. Phys. Lett. \textbf{97}, 043508 (2010).


\bibitem{Wagner_10}
M. Wagner, H. Schneider, D. Stehr, S. Winnerl, A. M. Andrews, S. Schartner and G. Strasser, M. Helm, Phys. Rev. Lett. \textbf{105}, 167401 (2010).

\bibitem{Teich_13}
M. Teich, M. Wagner1, H. Schneider and M. Helm1, New J. Phys. \textbf{15},065007 (2013).

\bibitem{Morina_15}
S. Morina, O. V. Kibis, A. A. Pervishko and I. A. Shelykh
Phys. Rev. B \textbf{91}, 155312 (2015).

\bibitem{Pervishko_15}
A. A. Pervishko, O. V. Kibis, S. Morina and I. A. Shelykh
Phys. Rev. B \textbf{92}, 205403 (2015).

\bibitem{Dini_16}
K. Dini, O. V. Kibis and I. A. Shelykh
Phys. Rev. B \textbf{93}, 235411 (2016).


\bibitem{Monroy_16}
R. Vega Monroy and G. Salazar Cohen, 
Nano Lett. \textbf{16}, 6797–6801 (2016).

\bibitem{Kibis_16}
O. V. Kibis, S. Morina, K. Dini and I. A. Shelykh,
Phys. Rev. B \textbf{93}, 115420 (2016).

\bibitem{Kristinsson_16}
K. Kristinsson, O. V. Kibis, S. Morina and I. A. Shelykh, 
Sci. Rep. \textbf{6}, 20082 (2016).
 
\bibitem{Iorsh_17}
I. V. Iorsh, K. Dini, O. V. Kibis and I. A. Shelykh,
Phys. Rev. B \textbf{96}, 155432 (2017).
 
\bibitem{DalLago_17}
V. Dal Lago, E. Suárez Morell and L. E. F. Foa Torres,
Phys. Rev. B \textbf{96}, 235409 (2017).

\bibitem{Oka_09}
T. Oka, H. Aoki,
Phys. Rev. B \textbf{79} and 081406 (2009).

\bibitem{Usaj_14}
G. Usaj, P. M. Perez-Piskunow, L. E. F. Foa Torres and C. A. Balseiro
Phys. Rev. B \textbf{90}, 115423 (2014).

\bibitem{Torres_14}
L.E.F Foa Torres, P.M Perez-Piskunow, C.A. Balseiro and G. Usaj,
Phys. Rev. Lett. \textbf{113}, 266801 (2014).

\bibitem{Yudin_16}
D Yudin, O. V. Kibis and I A Shelykh,
New J. Phys. \textbf{18}, 103014 (2016).

\bibitem{Kibis_17}
O. V. Kibis, K. Dini, I. V. Iorsh and I. A. Shelykh,
Phys. Rev. B \textbf{95}, 125401 (2017).

\bibitem{Morina_17}
S Morina, K Dini, IV Iorsh and IA Shelykh
arXiv preprint arXiv:1711.07313 (2017).

\bibitem{Hanggi_98}
P. H\"anngi, Driven quantum systems. In {\it Quantum Transport and
Dissipation} T. Dittrich, P. Hanggi, G. L. Ingold, B. Kramer and G. Sch\"on, W. Zwerger, Eds. Wiley and Weinheim, 1998.

\bibitem{Kohler_05}
S. Kohler, J. Lehmann and P. Hänggi, Phys. Rep. {\bf 406}, 379--446 (2005).

\bibitem{Bukov_15} 
M. Bukov, L. D'Alessio, A. Polkovnikov, Adv. Phys. \textbf{64}, 139--226 (2015).

\bibitem{Goldman_14}
N. Goldman and J. Dalibard,
Phys. Rev. X \textbf{4}, 031027 (2014).

\bibitem{Rahav_03}
S. Rahav, I Gilary, and S. Fishman
Phys. Rev. A \textbf{68}, 013820 (2003).

\bibitem{Eckardt_15}
A. Eckardt and E. Anisimovas,
New J. Phys.  \textbf{17}, 093039 (2015).

\bibitem{Sachs_11}
B. Sachs, T. O. Wehling, M. I. Katsnelson, and A. I. Lichtenstein
Phys. Rev. B \textbf{84}, 195414 (2011).

\bibitem{Jung_15}
J. Jung, A. M. DaSilva, A. H. MacDonald and S. Adam
Nat. Com. \textbf{6}, Article number: 6308 (2015).

\end{thebibliography}
\end{document}